# Multispectral Electro-Optical Surfaces: from Visible to Microwave

M. Said Ergoktas[1,2], Gokhan Bakan[1,2], Evgeniya Kovalska[3], Lewis W. Le Fevre[2,6], Richard P. Fields[2], Pietro Steiner[1,2], Xiaoxiao Yu[1,2], Omer Salihoglu[3], Sinan Balci[4], Kostya Novoselov[2,5], Robert A. W. Dryfe[2,6,7], Coskun Kocabas[1,2,7,*]

1. Department of Materials, University of Manchester, Manchester, M13 9PL, UK
2. National Graphene Institute, University of Manchester, Manchester, M13 9PL, UK
3. Department of Physics, Bilkent University, Ankara, Turkey
4. Department of Photonics, Izmir Institute of Technology, Izmir, Turkey
5. Department of Physics and Astronomy, University of Manchester, Manchester, M13 9PL, UK
6. Department of Chemistry, University of Manchester, Manchester, M13 9PL, UK
7. Henry Royce Institute for Advanced Materials, University of Manchester, Manchester, M13 9PL, UK

[*]coskun.kocabas@manchester.ac.uk

**Optical materials with colour-changing abilities have been explored for display devices[1], smart windows[2], or modulation of visual appearance[3–5]. The efficiency of these materials, however, has strong wavelength dependence, which limits their functionality to a narrow spectral range. Here, we report graphene-based electro-optical devices with unprecedented optical tunability covering the entire electromagnetic spectrum from the visible to microwave. We achieve this non-volatile and reversible tunability by electro-intercalation of lithium into graphene layers in an optically accessible device structure. This unique colour-changing capability, together with area-selective intercalation, inspires fabrication of new multispectral devices, including display devices and electro-optical camouflage coating. We anticipate that these results provide realistic approaches for programmable smart optical surfaces with a potential utility in many scientific and engineering fields.**

Conventional optoelectronic devices are intentionally designed to operate at a certain wavelength range. There might be new opportunities for optoelectronic devices that can operate over seemingly unrelated wavelengths. Applications that could exploit these multispectral operations include, for example, adaptive camouflage coatings that can perform concealment both in the visible and infrared wavelengths, a display device that can encode information in different wavelengths or a dynamic thermal blanket that can selectively reflect visible or infrared light. Such multispectral devices could be possible, if they can overcome existing challenges of requiring (1) broadband electro-optical tunability, (2) multispectral device structure, and (3) fast and non-volatile switching. The research on tunable optical materials has primarily focused on controlling the state of materials to alter the optical response. Phase-change[6–8] and electrochromic[9,10] materials are two examples that are capable of colour change triggered by temperature change or electric field. The existing device structures such as the requirement of a conductive top electrode limit their spectral operation range and utility for a multispectral application.[11,12]

Electro-intercalation of atomic or molecular ions into layered materials can rise to the challenge by providing the host material new means of controlling its electrical, thermal, and magnetic properties.[13,14] Especially graphite intercalation compounds (GICs) permit broad variations in these physical parameters.[15,16] Tuning optical properties of the host material *via* intercalation, however, has been overlooked mainly due to the difficulties of integration of optical devices with electrochemical cells.[16,17] In this Letter, we report an electrochemical optical platform with non-volatile and reversible reflectivity modulation covering the entire electromagnetic spectrum from the visible to the microwave regime. This platform differs from previously reported graphene devices, which are designed primarily for specific wavelength ranges of microwave,[18] THz,[19] infrared,[4,5] and visible[20] using single and multilayer graphene. Overcoming the challenge of extending the coverage to the visible while keeping the optical activity at the longer wavelengths has led to the innovative device structure reported here to reach the ultimate multispectral operation.

**Figure 1**a illustrates the structure of the device and its operation principle. The device takes advantage of multilayer graphene (MLG) as the anode and as well as the optically active material. High electrical conductivity of MLG (<50 Ω/sq) eliminates the requirement of an additional top electrode layer. The key feature of the device is that MLG is optically accessible under a transparent protection layer. This is achieved by laminating MLG and vacuum sealing the device in a low-density polyethylene pouch that has very high optical transparency from the visible to microwave wavelengths (**Figure S1**). The airtight packaging is crucial for a stable non-volatile operation. Aluminium foil, coated with lithium-doped nickel manganese cobalt oxide (NMC), is used as the cathode and the source of $Li^+$ ions. Electro-intercalation of $Li^+$ ions into MLG (charging the device) is achieved by applying a constant electric current (~1 $mA/cm^2$) towards the cathode (**Figure 1**b). The device voltage was monitored to track the state of charge. Reversing the current direction, or simply shorting through an electrical load (**Figure S2** and Video S1), de-intercalates the ions from MLG, making the operation reversible and repeatable.[21] During intercalation (or de-intercalation), the electrical and optical properties of MLG change drastically. Discharged device appears dark grey owing to the high absorptivity of the top graphene layer. When the device is fully charged (at ~3.8 V), the graphene layer appears gold in colour (**Figure 1**c), similar to stage 1 GIC.[15,22] Besides the visible colour change, the infrared properties, following a similar trend, exhibit high absorptivity (hence high emissivity) when discharged and high reflectivity (low emissivity) when charged (**Figure 1**d). Thermal imaging of the device was performed outdoors to minimize the ambient thermal emission reflected from the device. The charged device appears significantly colder to the thermal camera compared to its actual temperature owing to its low infrared emissivity.

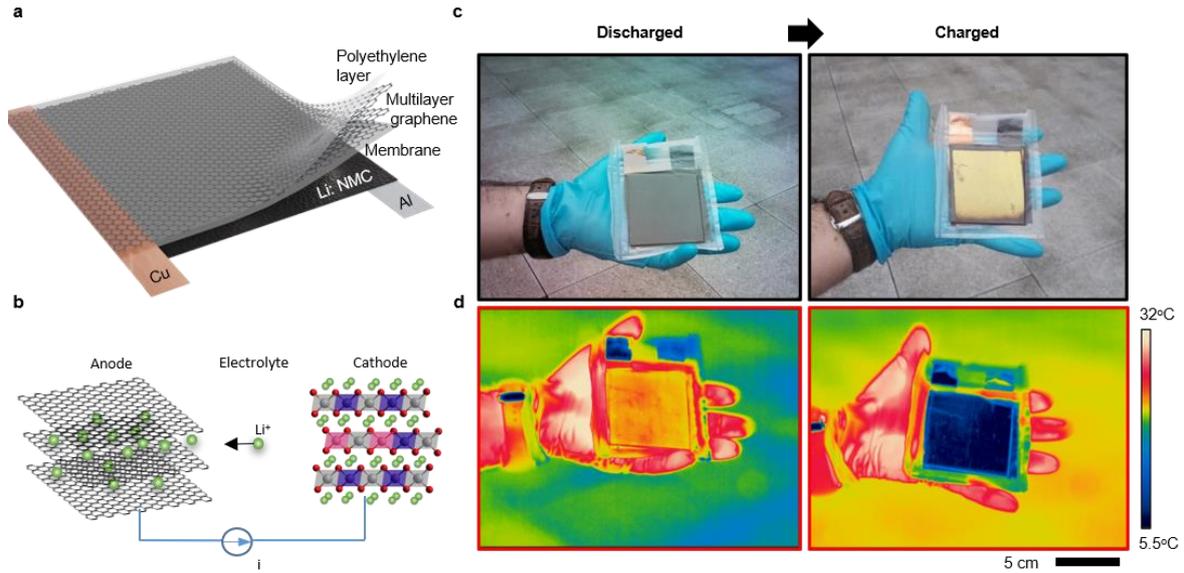

**Figure 1. Device structure and operation principle. a,** Schematic drawing of the device consisting of polyethylene protection, multilayer graphene (anode), electrolyte-containing membrane, and aluminium foil coated with lithium-doped Ni-Mn-Co-oxide (cathode). **b,** Illustration of Li$^+$ ion intercalation into graphene layers. **c,** Visible and **d,** Infrared images of the device at fully discharged and charged states showing non-volatile multispectral tunability.

To examine the multispectral response, we measured the reflection spectra of the devices at different doping levels. By combining the reflection measurements from three different spectrometers (Ultraviolet/Visible/Near-infrared, Fourier Transform Infrared (FTIR) Spectrometer, and Time-domain THz Spectrometer), we obtained the reflectivity change from the ultraviolet (300 nm of wavelength) to microwave (3 mm of wavelength) as shown in **Figure 2**a. The reflectivity can be modulated substantially in the entire range of the non-ionising part of the electromagnetic spectrum. **Figure 2**b compares the reflectivity modulation for three different wavelengths that are orders of magnitude apart. The modulation onset varies with the wavelength of light. THz reflectivity increases at the very early stage of doping. On the other hand, modulation in the visible regime appears at higher doping levels. The sharp drop in the reflectance is an indicator of the free-carrier plasma resonance, $\omega_p$. This reflection minimum shifts from the far-infrared to ultraviolet for increasing doping. We obtained the plasma frequency, shown in **Figure 2**c, by fitting the Drude and interband models to the experimental reflection spectra: $\varepsilon = \varepsilon_\infty + \varepsilon_{int} + i\frac{\omega_p^2 \tau}{\omega(1-i\omega\tau)}$, where $\varepsilon_\infty$ is the background dielectric constant, $\varepsilon_{int}$ accounts for the contribution of interband transitions, $\tau$ is the relaxation time of carriers, and $\omega$ is the angular frequency of light (see Supplementary Note 1 and **Figure S3**). For the same device structure, **Figure 2**d shows the variation in the sheet resistance of MLG and the device voltage during charging cycle. Clear steps are observed in these measurements indicating the distinct intercalation stages. We also performed *in-situ* Raman measurements during charging (acquired by 457 nm laser excitation) (**Figure S4**). As the device charges, the G-band in the Raman spectrum at 1580 cm$^{-1}$ first shifts to 1600 cm$^{-1}$ and then disappears. These observations are attributed to increasing the C-C bond force constants, highly conductive GICs,[23] and Pauli blocking of electron-phonon coupling due to the doping of Li$^+$ ions.[24]

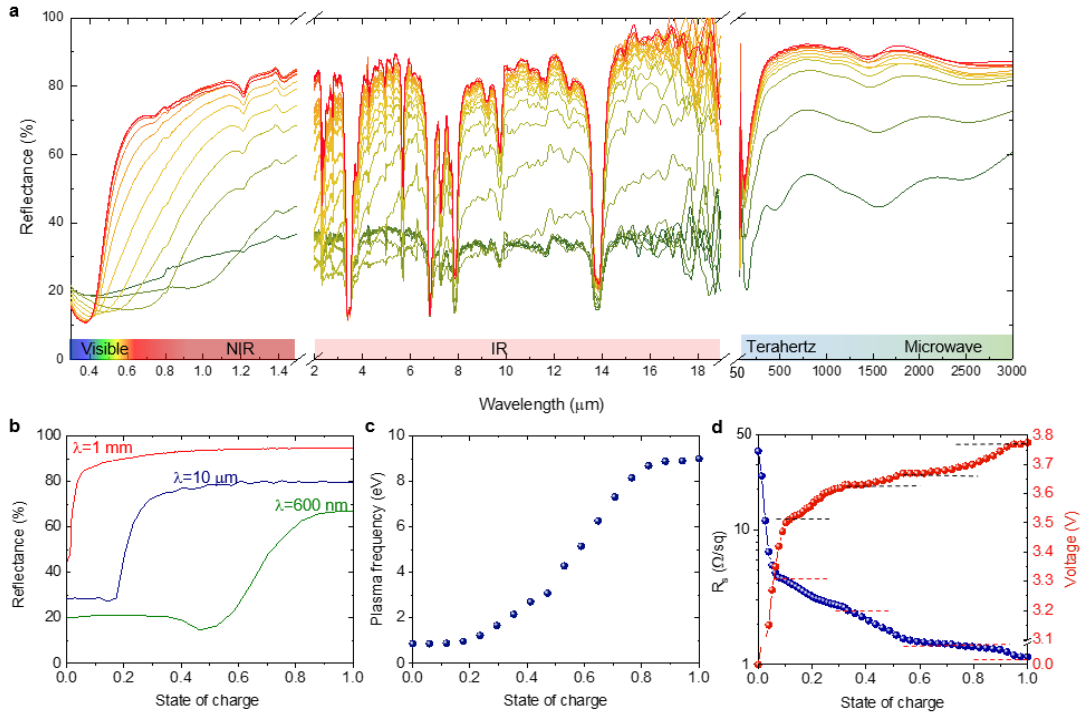

**Figure 2. Multispectral tunability from ultraviolet to microwave.** **a,** Measured reflection spectra covering the visible, infrared, THz, and microwave regimes for different doping levels (state of charge). The general trend is increasing reflectance with increasing state of charge. The sharp reflection dips in the infrared spectrum are due to absorption by the top polyethylene protection layer. **b,** Variation in the reflectivity of the device at the selected wavelengths of 600 nm, 10 μm, and 1 mm. **c,** Change in the plasma frequency as a function of the state of charge. **d,** Device voltage and sheet resistance of MLG during charging. Dashed lines mark the different stages of lithiation of graphene.

To show the promises of the tunable multispectral optical surfaces, we demonstrate two proof-of-concept (PoC) devices. The first device is a display capable of showing images in the visible, infrared, and THz wavelengths, simultaneously. **Figure 3**a shows cross-section illustration of the device, which uses continuous graphene layer and area-selective intercalation of $Li^+$ ions enabled by patterning the cathode. Controlling the voltage applied to an individual electrode enables varying the pixel's spectral response. **Figure 3**b showcases the display device with its back electrode patterned into letters "N", "G", and "I" (the initials of National Graphene Institute). Multispectral imaging of the device was performed by a CMOS camera operating in the visible regime (400 – 700 nm), a long-wave infrared camera (8 – 13 µm), and THz raster-scan imaging system (0.5 – 2 THz). Applying 3.8, 3.5, and 3.3 V to "N", "G", and "I" pixels, respectively, revealed "N" in all wavelengths, "G" in the infrared and THz, and "I" only in the THz. Such an example is first of its kind and, beyond the demonstrated function as multispectral display, it has the potential to be used for optical security applications in which an encoded message can only be revealed with the correct key that corresponds to applying a specific voltage to the device and imaging in a specific range of wavelengths for our technology. The demonstrated concept relies on non-overlapping voltage ranges for different spectral regimes, e.g., being able to change the infrared emission while maintaining the visible response. This feature is visualized on a device by (i) changing both the visible and infrared response, (ii) changing only the infrared response while the visible reflection is always low, and (iii) changing only the visible response while the infrared emission is always low (Video S2). The devices leverage flexible printed circuit board technology to build flexible, multipixel, and multispectral displays as demonstrated in **Figure S5**.

The second PoC device is a multispectral adaptive camouflage system that has a unique capability of modifying its visible and infrared appearance dynamically. The adaptive visibility was achieved by patterning the back electrode into a camouflage pattern. Then, specific parts of the device were activated to match its appearance to that of a spatially varying background (**Figure 3**c). The graph in **Figure 3**c shows time trace of the red, green, and blue (RGB) colour indices during adaptation. Similarly, the device can adapt its thermal appearance. **Figure 3**d, e shows thermal images of a device as it adapts its infrared emission to cold and hot backgrounds (see Video S3). These demonstrations highlight a comprehensive case of thermal camouflage where the background infrared emission dynamically changes taking into account moving targets and changing environmental factors, e.g., ambient temperature, clouds (see Supplementary Note 2 for further discussions). While the background in these examples was kept spatially uniform, we demonstrated that the approach used for the visible camouflage can be adapted for a spatially varying thermal background (**Figure S8**). The time-trace of the infrared camouflage process reveals the high transition speed and quantifies how well the infrared emission from the device (rendered as the apparent temperature by the camera) adapts to the background (highlighted areas in **Figure 3**d, e). The camouflage operation can easily be extended into the mid-wave (3 – 5 µm) and short-wave (1.4 – 3 µm) infrared regimes as the emissivity modulation in the entire infrared spectrum is high and relatively uniform (**Figure S9**).

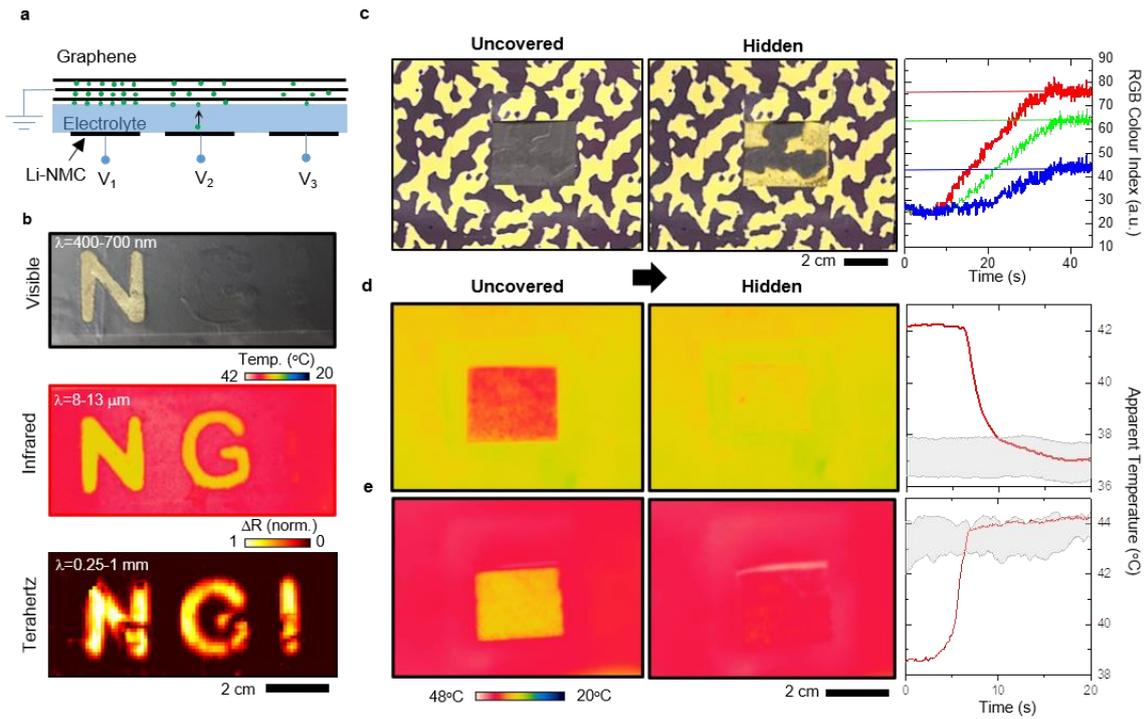

**Figure 3. Multispectral display and adaptive camouflage. a,** Cross section illustration of PoC multispectral display based on area-selective lithiation. Patterned cathodes define individually addressable pixels. The state of charge determines the operation wavelength. **b,** Images of a display device rendered in the visible, infrared, and THz regimes. **c,** Demonstration of adaptive visible camouflage. The device blends into a spatially varying background by matching its visible appearance. The graph shows the change in red, green, and blue (RGB) colour indices from the active region of the device as it blends in. Vertical lines mark the background RGB indices. **d, e,** Demonstration of adaptive infrared camouflage. The device can adapt its infrared emission to blend in cold (**d**) and hot (**e**) backgrounds. Graphs show the time-trace of the apparent device temperature during adaptation. The shaded areas show the background temperature range.

In the final part, we discuss how the multispectral tunability can improve adaptive thermal management of satellites in outer space by balancing the absorbed solar radiation (wavelength range: 0.3 – 2.5 µm) and the internal heat generated by satellite electronics with the emitted infrared radiation (peaks ~10 µm) (**Figure 4**a-c). Maintaining this balance with passive optical components while the satellite orbits the earth creates major design challenges especially for small satellites. The demonstrated multispectral electro-optical tunability in the visible and infrared regimes can address this need by varying the visible absorptivity and infrared emissivity with the state of charge (**Figure 4**d). For the calculations, we used a low orbit (altitude of 600 km) CubeSat as a model system for two extreme cases: (i) when the satellite is exposed to direct sunlight, (ii) in earth's shadow. For the first case, rejecting the incident solar irradiance is the priority as the solar irradiance at the satellite's orbit (~1347 W/m$^2$) is much larger than a blackbody at 30 °C can emit to the deep space (~479 W/m$^2$). Varying the state of charge on surfaces that are not exposed to sunlight provides a dynamic net cooling rate as large as 141 W/m$^2$. For the second case, all sides but the earth-facing one are exposed to the deep space and the satellite receives no direct or indirect solar radiation. The net cooling rate in this case is evaluated between 95 and 282 W/m$^2$ with varying state of charge (**Figure 4**d). In contrast, a thermal coating with static optical properties can only provide a static cooling rate making the satellite prone to larger temperature fluctuations with varying satellite position and activity (see Supplementary Note 3 for a detailed analysis).

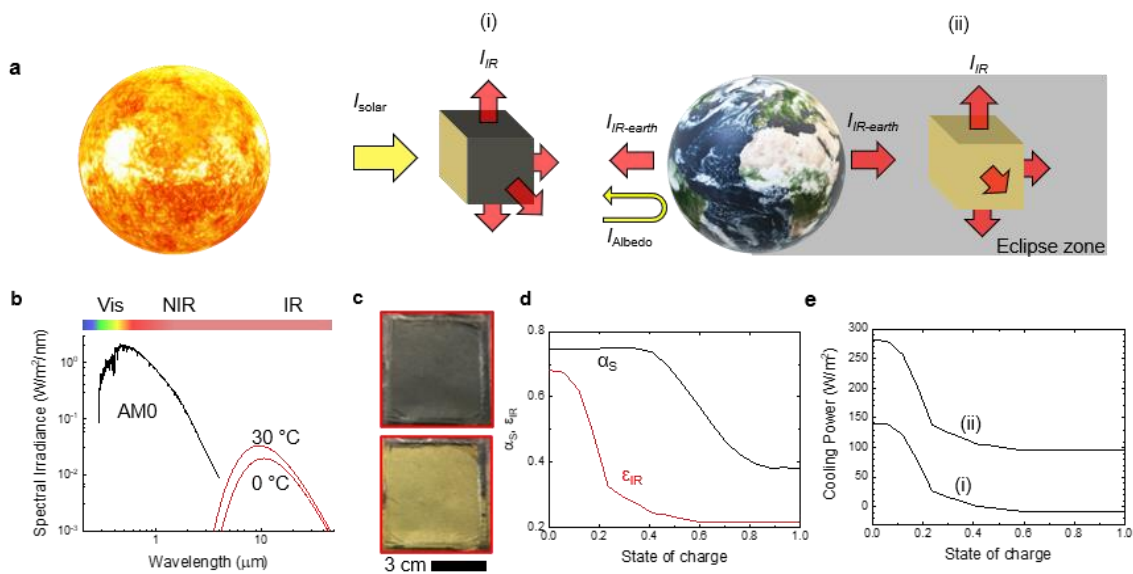

**Figure 4. Adaptive thermal management for satellites**. **a,** Schematic drawing of a CubeSat orbiting the earth, illustrating the solar and infrared irradiance on it. **b,** Spectral solar irradiance on a low-earth-orbiting satellite and blackbody radiation at 0 and 30 °C. **c,** Photographs of a large-area device at zero and full state of charge. This device can fully cover single side of a CubeSat. **d,** Weighted solar absorptivity and infrared emissivity of the developed multispectral surfaces as a function of state of charge. **e,** Calculated cooling power of the satellite (i) under direct sunlight and (ii) in earth's shadow.

The temporal response of the device is of critical importance for these applications. The switching speed is limited by the intercalation process, quantified as the state of charge in this work that has linear dependence on the device current. For a moderate current density of ~1mA/cm$^2$, full charging and accompanying optical changes take place in 20 s being superior to the electrochromic alternatives operating in the visible regime with switching durations ranging from tens of seconds to minutes.[9] Switching the infrared response is much faster in comparison with less than 1 s response time (**Figure S10**).

The long-term steady operation of the devices necessitates mitigating the high volatility of the electrode. The same problem present for Li-ion batteries is solved by airtight packing and solid-electrolyte interface.[25] Following this example, our devices were vacuum sealed in polyethylene pouches to press all layers against each other while keeping the graphene layers optically accessible. The devices can endure long duration on/off cycles without significant degradation to the electrical and optical performance (**Figure S10** shows 415 cycles of switching the infrared emission on and off performed in ~30 minutes). Beyond the PoC demonstrations included here, the developed technology inspires new applications such as battery level indicator for Li-ion batteries (**Figure S11**).

In conclusion, we demonstrate multispectral electro-optical devices operating over the entire electromagnetic spectrum from the visible to microwave. This unprecedented capability would inspire new technologies that has a very high potential to reshape the tunable optics either as a standalone unit or by incorporating with established light manipulation approaches. Furthermore, the broad electro-optical tunability of the plasma frequency of the multilayer graphene, yields a unique electrically tuneable plasmonic system which can open multispectral active plasmonics.

**Acknowledgements:** This research is supported by European Research Council through ERC-Consolidator Grant (grant no 682723, SmartGraphene). In addition, we acknowledge Graphene Engineering Innovation Centre (GEIC) for access to the CVD system.

**Author contributions:** C.K. conceived the idea. M.S.E. synthesized the graphene samples and fabricated the devices. M.S.E., G.B., E.K., O.S., S.B., and C.K. performed the experiments. L.F., R.F., P.S., and X.Y. helped with the fabrication of the devices. G.B., M.S.E., and C.K. analysed the data and wrote the manuscript with input from all the authors. All authors discussed the results and contributed to the scientific interpretation as well as to the writing of the manuscript.

**Additional information:** C.K. is involved activities towards commercialization of graphene based optical surfaces under SmartIR Ltd. The work is subject to patent application by C.K., M.S.E., and G.B. The remaining authors declare no competing financial interests.